\documentclass[aps,prl,twocolumn,superscriptaddress,letterpaper,longbibliography]{revtex4-2}
\usepackage{graphicx}
\usepackage{bm}
\usepackage{amssymb}
\usepackage{color}
\usepackage{amsmath}
\usepackage{ulem}
\usepackage{bbold}
\usepackage[mathlines]{lineno}
\usepackage[colorlinks=True,citecolor=blue,linkcolor=blue,allcolors=blue]{hyperref}
\usepackage{enumitem}
\usepackage[utf8]{inputenc}

\begin{document}

\title{Symmetry Analysis of Anomalous Floquet Topological Phases}
\author{Weiwei Zhu}
\affiliation{Department of Physics, National University of Singapore, Singapore 117542, Singapore}

\author{Yidong Chong}
\email{yidong@ntu.edu.sg}
\affiliation{Division of Physics and Applied Physics, School of Physical and Mathematical Sciences, Nanyang Technological University, Singapore 637371, Singapore}
\affiliation{Centre for Disruptive Photonic Technologies, Nanyang Technological University, Singapore 637371, Singapore}
\author{Jiangbin Gong}
\email{phygj@nus.edu.sg}
\affiliation{Department of Physics, National University of Singapore, Singapore 117542, Singapore}

\begin{abstract}
The topological characterization of  nonequilibrium topological matter is highly nontrivial because familiar approaches designed for equilibrium topological phases may not apply. In the presence of crystal symmetry, Floquet topological insulator states cannot be easily distinguished from normal insulators by a set of symmetry eigenvalues at high symmetry points in the  Brillouin zone.   This work advocates a physically motivated, easy-to-implement approach to enhance the symmetry analysis to distinguish between a variety of Floquet topological phases.  Using a two-dimensional inversion-symmetric periodically-driven system as an example, we show that the symmetry eigenvalues for anomalous Floquet topological states, of both first-order and second-order, are the same as for normal atomic insulators. However, the topological states can be distinguished from one another and from normal insulators by inspecting the occurrence of stable symmetry inversion points in their microscopic dynamics.   The analysis points to a simple picture for understanding how topological boundary states can coexist with localized bulk states in anomalous Floquet topological phases.
\end{abstract}

\maketitle
{\it Introduction.}---Topological insulators (TIs) are phases of matter exhibiting robust gapless boundary states \cite{Hasan2010,Qi2011}, typically tied to a nonzero topological band invariant defined in momentum space \cite{Zak1985,Fu2007,Fang2012}.  Recently, the development of the theory of TIs with crystal symmetry has provided a fruitful position-space picture for analyzing and classifying TIs \cite{Zak1981, Soluyanov2011, Slager2013, Read2017, Kruthoff2017, Po2017, Bradlyn2017, Khalaf2018, Cano2018a, Cano2018, Po2020, Cano2020}. Specifically, TIs rule out symmetric and exponentially localized Wannier functions,  a concept called Wannier obstruction.  Given a crystal symmetry, the band representations for all possible atomic insulators at highly symmetric momentum points (HSMPs) can be obtained by placing different orbitals at maximal Wyckoff positions; TIs, however, cannot be expressed in this way~\cite{Po2020,Cano2020}.  This offers a way to identify TI phases in real materials~\cite{Song2018,Tang2019a,Tang2019,Zhang2019,Vergniory2019}, and has stimulated the discovery of new topological phases such as higher-order topological insulators~\cite{Schindler2018,You2018,Benalcazar2019} and fragile topological insulators~\cite{Po2018,Bradlyn2019,Hwang2019,Bouhon2019,Song2020a,Liu2019,Song2020}.

Floquet TIs are a special class of topological phases that arise in periodically driven systems \cite{Kitagawa2010, Gong2012, Rechtsman2013,Platero2013,Kundu2013,zhao2013, Hubener2017, Rudner2020, McIver2020, Wintersperger2020}, possessing numerous features not found in equilibrium systems.  These include $\pi$ modes~\cite{Jiang2011,Tong2013,Bomantara2019,Gong2018,Gong2020}, anomalous Floquet topological insulators (AFIs)~\cite{Rudner2014, Liang2013, Nathan2015, Hu2015, Leykam2016, Delplace2017,longwen2018, Wintersperger2020}, and anomalous Floquet higher-order topological insulators (AFHOTIs)~\cite{Peng2019, Rodriguez-Vega2019, Hu2020, Huang2020, Zhu2021, Zhu2020, Zhang2020}.
Symmetry-based analyses of some Floquet topological phases where Wannier obstruction still works, such as Floquet fragile topological phases and Floquet Chern insulators~\cite{Zhang2020a}, can still be performed in a similar manner as in their equilibrium counterparts.
However, in general, the very existence of AFIs indicates that Wannier obstruction alone ceases to be a useful tool for characterizing topological phases, when it comes to nonequilibrium systems.
  How to distinguish between anomalous Floquet topological phases by symmetry analysis is still an open question.

In this work, we introduce a symmetry-based characterization method that is specifically geared toward Floquet systems, which treats different types of Floquet TIs on the same footing.  This is based on considering the microscopic dynamics of Floquet topological systems in connection with ``symmetry inversion points'' that change along the time dimension.  We demonstrate this approach via the symmetry analysis of AFI and AFHOTI phases in periodically-driven bipartite two-dimensional (2D) lattices with inversion symmetry.  The symmetry eigenvalues at HSMPs for AFIs and AFHOTIs are the same as atomic insulators with Wyckoff position at $1a$ (the center of the unit cell). Yet even though these phases have the same Wannier representation as atomic insulators, they cannot be deformed into known normal insulators
due to the existence of stable singularities characterized by symmetry inversion points in the time dimension.  Remarkably, it is shown that these symmetry inversion points can be analyzed by fully exploiting the symmetries of the lattice.  Finally, we exploit a mapping to an oriented scattering network to qualitatively explain why AFIs and AFHOTIs with localized bulk states can nonetheless accommodate topological boundary states.

\begin{figure}
\includegraphics[width=1\linewidth]{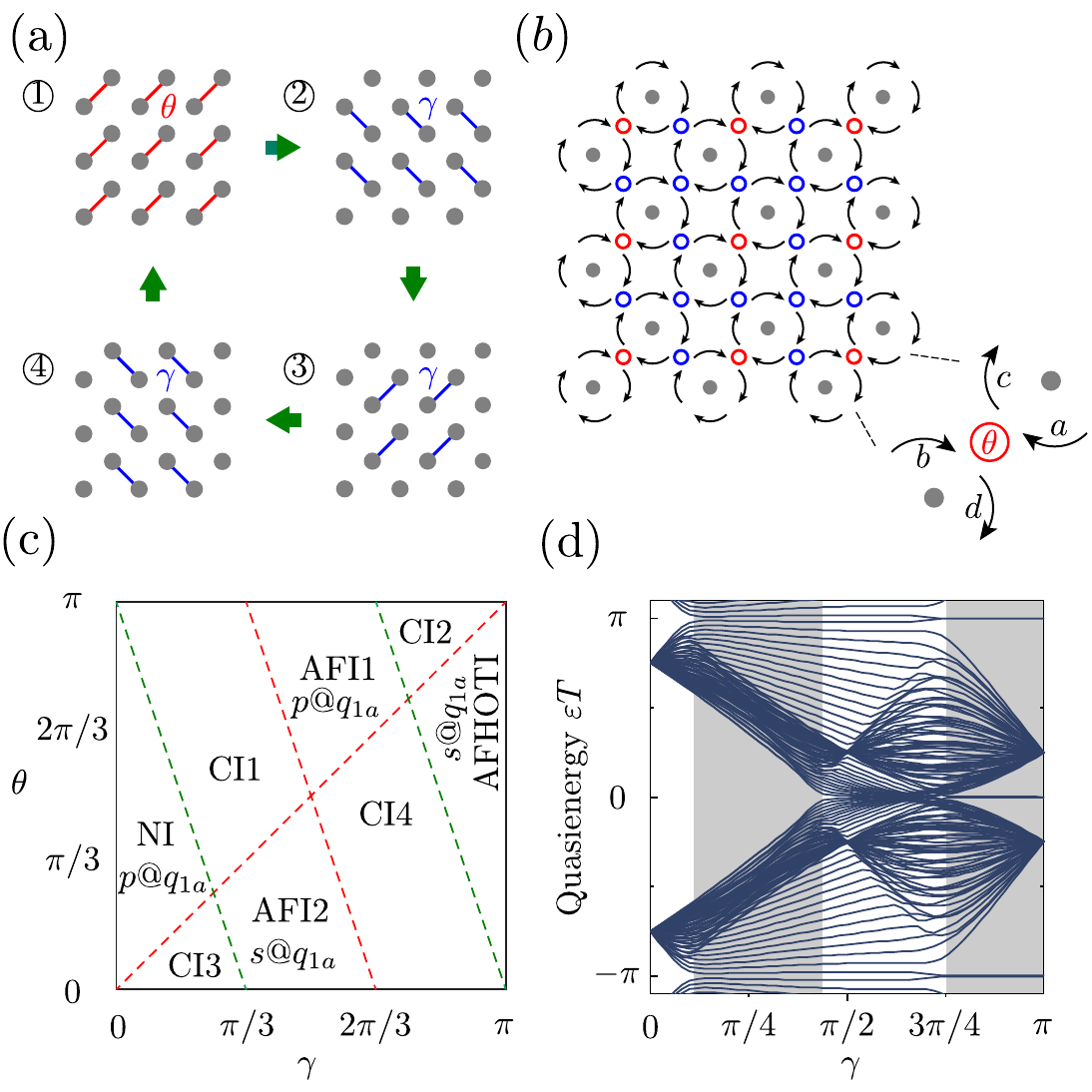}
\caption{ Model and phase diagram. (a) Periodically driving bipartite lattice. The model is composed of four time steps. Step $1$ consists of intra-cell couplings $\theta$, while steps $2$--$4$ consist of inter-cell couplings $\gamma$. (b) The periodically driving bipartite lattice expressed as a directed scattering network, with red (blue) circles representing coupling $\theta$ ($\gamma$).  (c) The phase diagram of the system, containing various topological phases including one NI, two AFIs, four CIs and one AFHOTI. The Wannier representations for the NI, AFIs, and AFHOTI are marked; they are $s$ or $p$ orbitals at position $1a$. (d) Spectrum of a finite structure as a function of $\gamma$, with $\theta=3\pi/4$.}
\label{model}
\end{figure}

{\it Model.}---Consider a two-dimensional periodically driven bipartite lattice as shown in Fig.~\ref{model}(a), which was recently proposed by us to realize AFHOTIs~\cite{Zhu2021,Zhu2020}. The model is composed of four steps and in each step the lattice is fully dimerized. Step $1$ only contains intra-cell couplings $\theta$, whereas steps $2$--$4$ contain inter-cell couplings $\gamma$.  The associated Hamiltonian is
\begin{eqnarray}
 H(\mathbf{k},t)=\left\{
\begin{array}{cc}
H_{1}(\mathbf{k})&0<t\leq T/4\;,\\
H_2(\mathbf{k})&T/4<t\leq T/2\;,\\
H_3(\mathbf{k})&T/2<t\leq 3T/4\;,\\
H_4(\mathbf{k})&3T/4<t\leq T\;,
\end{array}
\right.
\label{eq1}
\end{eqnarray}
where
\begin{equation}\label{eq2}
 H_{1}(\mathbf{k})=\theta\sigma_x
\end{equation}
and
\begin{equation}
H_{m}(\mathbf{k})=\gamma(e^{ib_m\cdot\mathbf{k}}\sigma^{+}+h.c.),
\label{eq3}
\end{equation}
for $m=2,3,4$.  Here, $\theta$ is a coupling parameter, $\sigma^{\pm}=(\sigma_{x}\pm i\sigma_{y})/2$ where $\sigma_{x,y,z}$ are Pauli matrices; and the vectors $\mathbf{b}_{m}$ are given by $\mathbf{b}_{2}=(0,a)$, $\mathbf{b}_{3}=(a,a)$ and $\mathbf{b}_{4}=(a,0)$, where $a$ is the lattice constant. Throughout we adopt a unit system where $\hbar=1$ and set $T=4$.  It will also be useful to note that this model is equivalent to the oriented scattering network model shown in Fig.~\ref{model}(b), where the red (blue) circle representing intra-cell couplings $\theta$ (inter-cell couplings $\gamma$).  The intra-cell scattering process is $[c,~d]^T = S(\theta)~[a,~b]^T$, where
\begin{equation}
  S(\theta)\equiv\mathcal{T}\mathrm{exp}\big[-i\int_{0}^{T/4} H(\mathbf{k},\tau)\,d\tau\big]=\exp(-i\theta\sigma_x)
\end{equation}
is the time evolution operator generated by step 1 in Fig.~\ref{model}(a) ($\mathcal{T}$ denotes the time-ordering operator). We seek quasi-stationary solutions that have the form $|\psi(\mathbf{k},t)\rangle=e^{-i\varepsilon(\mathbf{k}) t}|\phi(\mathbf{k},t)\rangle$, where $|\phi(\mathbf{k},t)\rangle$ is periodic in time with period $T$.  The quasienergy band $\varepsilon(\mathbf{k})$ is then determined by the Floquet equation
\begin{equation}\label{eq4}
  U_T(\mathbf{k})\, |\phi(\mathbf{k})\rangle = e^{-i\varepsilon(\mathbf{k}) T} |\phi(\mathbf{k})\rangle\;,
\end{equation}
where $U_T(\mathbf{k}) \equiv \mathcal{T}\mathrm{exp}\big[\!-\!i\!\int_{0}^{T} H(\mathbf{k},\tau)\,d\tau\big]$ is the Floquet operator (i.e., the time evolution operator over one period).

Because the Hamiltonian in Eq.~(\ref{eq1}) is bipartite and obeys particle-hole symmetry $CH(\mathbf{k},t)C=-H^{*}(-\mathbf{k},t)$
(where $C=\sigma_z$), one expects only two possible quasienergy band gaps, one around $\varepsilon = 0$ and the other around $\varepsilon = \pi/T$.  Each gap may or may not support topological edge modes or corner modes.  The remarkably rich phase diagram of the model is shown in Fig.~\ref{model}(c), containing: one normal insulator (NI) phase, in which neither gap hosts topological boundary states; four different Chern insulator (CI) phases, in which only one gap hosts chiral edge modes; two anomalous Floquet topological insulator (AFI) phases, in which both gaps host chiral edge modes; and one anomalous Floquet higher-order topological insulator (AFHOTI) phase, in which both gaps host topological corner modes.  We will focus mainly on the NI, AFI and AFHOTI phases and their Wannier representations.

Fig.~\ref{model}(d) shows the quasienergy spectrum of a finite lattice, plotted against $\gamma$ for fixed $\theta=3\pi/4$.  This single band diagram exhibits, from left to right, the NI, CI, AFI and AFHOTI phases, consistent with the phase diagram of Fig.~\ref{model}(c).

\begin{figure}
\includegraphics[width=1\linewidth]{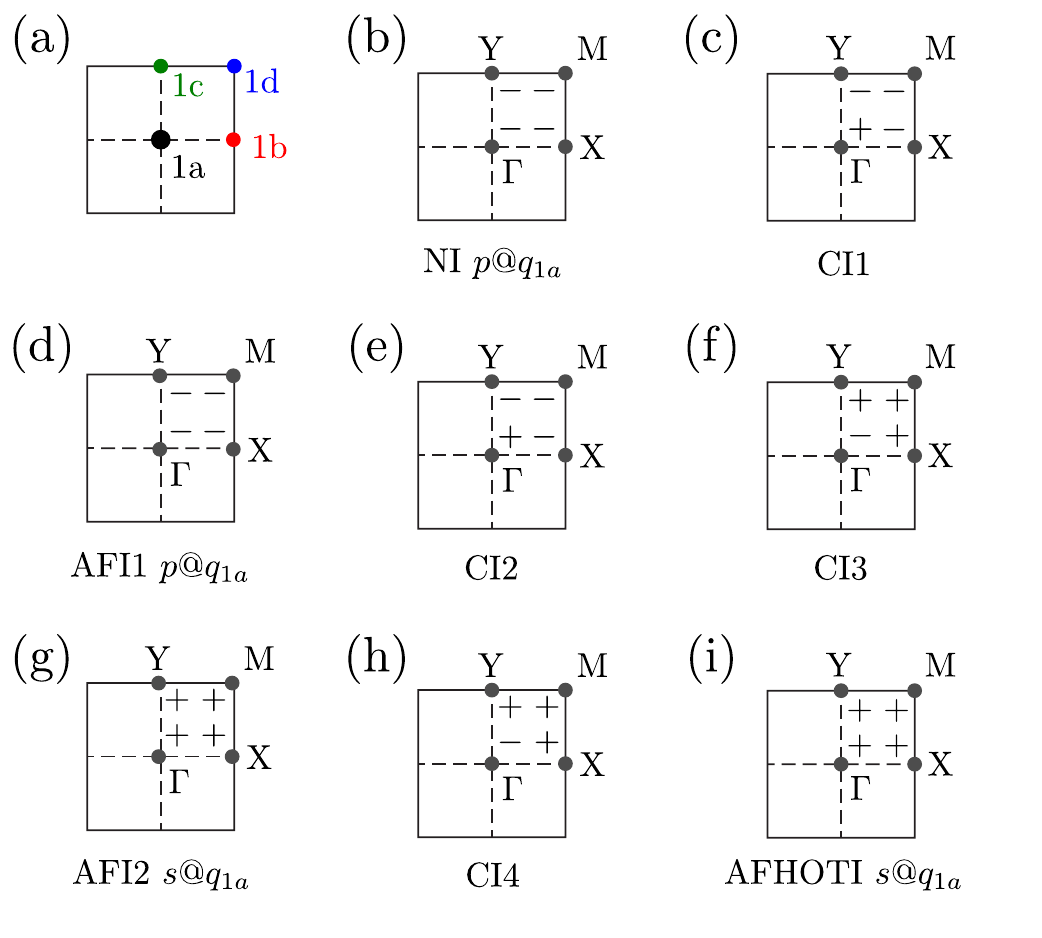}
\caption{Maximal Wyckoff positions and inversion symmetry eigenvalues. (a) Maximal Wyckoff positions for a square lattice with inversion symmetry. (b)-(i) Inversion symmetry eigenvalues at the highly symmetric momentum points $\Gamma$, $X$, $Y$, and $M$.  Each panel corresponds to one of the topological phases shown in Fig.~\ref{model}(c); the NI, AFI, and AFHOTI phases are labelled by their phase category.}
\label{Wyckoffposition}
\end{figure}

{\it Symmetry analysis and Wannier representation.}---The lattice described above obeys the inversion symmetry $\mathcal{I} H(\mathbf{k},t) \mathcal{I} = H(-\mathbf{k},t)$, where $\mathcal{I}=\sigma_x$. Using this symmetry, we can obtain the maximal Wyckoff positions, which are denoted as $1a$ (the center of the unit cell), $1b$, $1c$ and $1d$, as shown in Fig.~\ref{Wyckoffposition}(a).  The HSMPs occur at $\Gamma$, $X$, $Y$, and $M$, as shown in the other panels of Fig.~\ref{Wyckoffposition}. By placing an $s$ or $p$ orbital at the maximal Wyckoff position, the band representation at the HSMPs can be obtained, yielding the inversion symmetry eigenvalues shown in Table~\ref{table} for the maximal Wyckoff position at $1a$ (as shown later, only results at $1a$ are useful in our model, with the results  for maximal Wyckoff positions at $1b$, $1c$, and $1d$ shown in Supplementary Material).

\begin{table}
\caption{\label{table}Inversion symmetry eigenvalues at HSMPs for elementary band representation $s@q_{1a}$ and $p@q_{1a}$.}
\begin{ruledtabular}
\begin{tabular}{ccccc}
& $\,\Gamma\,$ & $\,$ X$\,$& $\,$Y$\,$ & $\,$M$\,$ \\
\hline
  $s@q_{1a}$ & $+1$ & $+1$ & $+1$ & $+1$ \\
  $p@q_{1a}$ & $-1$ & $-1$ & $-1$ & $-1$ \\
\end{tabular}
\end{ruledtabular}
\end{table}

Let us now examine the inversion symmetry eigenvalues of the various topological phases found in Fig.~\ref{model}(c). Due to the inversion symmetry, the Floquet operator at each HSMP satisfies $\mathcal{I} U_T \mathcal{I} = U_T$, where $U_T$ is evaluated at the HSMP in question. The inversion symmetry eigenvalues are then obtained by $\mathcal{I}|\phi\rangle=\pm|\phi\rangle$, where $|\phi\rangle$ is the lower-band Floquet eigenstate at the HSMP. The results are shown in Figs.~\ref{Wyckoffposition}(b)-(i).   The number of negative symmetry eigenvalues for the four CIs is either $1$ or $3$.  By contrast, atomic insulators have either $0$, $2$ or $4$ negative symmetry eigenvalues (see Supplemental Materials). This is unsurprising, since an odd number of negative symmetry eigenvalues implies a nonzero Chern number ($+1$ or $-1$ for the four CIs).

Furthermore, NI and AFI1 have exactly the same band representation as $p@q_{1a}$ in Table~\ref{table}, whereas AFI2 and AFHOTI share exactly the same band representation as $s@q_{1a}$, hinting that these phases have vanishing polarization and Chern number.  It thus appears that the nontrivial topological properties of the AFI and AFHOTI phases cannot be understood by Wannier obstruction.  One way to understand this is that the above band representation only captures the time evolution over one period, whereas anomalous Floquet topological phases are characterized not by Chern numbers but by a dynamical winding number derived from the time evolution \textit{within} one period~\cite{Rudner2014}.

{\it Stable symmetry inversion points in microscopic dynamics.}---Existing studies have shown that the AFI and AFHOTI phases may be characterized using singularities in the Floquet dynamics~\cite{Nathan2015, Hu2020, Zhang2020}.  Specifically, the time evolution within one period is captured by $U_t(\mathbf{k}) \equiv \mathcal{T}\mathrm{exp}\big[-i\int_{0}^{t} H(\mathbf{k},\tau)\,d\tau\big]$.  Consider the eigenvalue equation
\begin{equation}\label{eq5}
  U_t(\mathbf{k})\, |\phi(\mathbf{k},t)\rangle = e^{-i\varphi(\mathbf{k},t)} |\phi(\mathbf{k},t)\rangle\;,
\end{equation}
where $\varphi(\mathbf{k},t)$ forms phase bands in $(\mathbf{k},t)$ space (for $0\leq t\leq T$). The existence of singularities in $\varphi(\mathbf{k},t)$ forbids $U_T(\mathbf{k})$ from being continuously deformed into the Floquet evolution operator of an atomic insulator.  Unfortunately, the calculation of such phase bands is rather complicated in general, and one further needs to check the stability of the singularity by inspecting some Chern numbers surrounding the identified singularity point.   Instead, we will now show that by fully exploiting the crystal symmetry, it is possible to easily identify and analyze the singularities via symmetry inversion points (SIPs).

\begin{figure}
\includegraphics[width=1\linewidth]{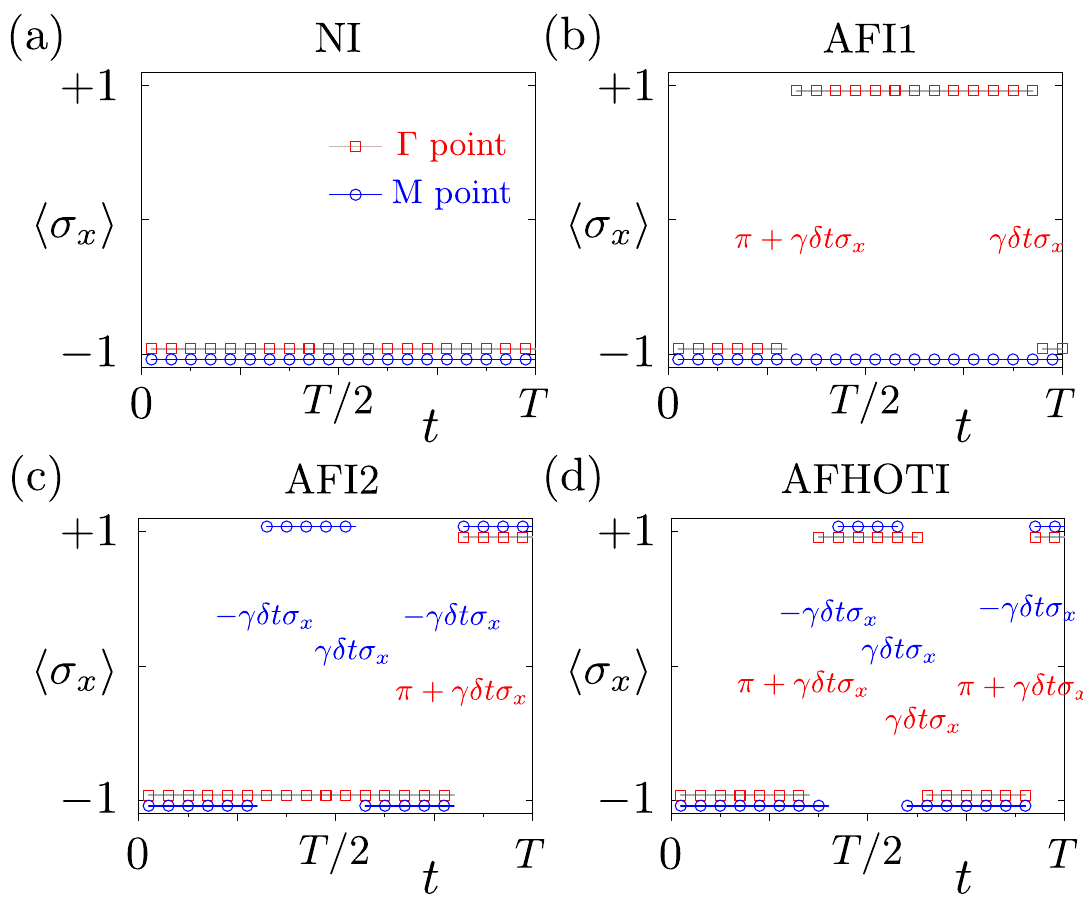}
\caption{SIPs for the different topological phases. (a)--(d) Plot of $\langle\sigma_x\rangle$ versus time within one period, calculated using Eq.~\eqref{eq6} for the NI, AFI1, AFI2 and AFHOTI phases.  Red (blue) data points correspond to results at $\Gamma$ ($M$).  The discontinuities correspond to SIPs, and the text labels indicate the effective continuous Hamiltonian near each SIP.  The lattice parameters are (a) $\theta=0.4\pi$, $\gamma=0.1\pi$,  (b) $\theta=0.9\pi$, $\gamma=0.4\pi$, (c) $\theta=0.1\pi$, $\gamma=0.4\pi$, and (d) $\theta=0.6\pi$, $\gamma=0.9\pi$.}
\label{symmetry}
\end{figure}

Consider a HSMP denoted by $\mathbf{K}$,  where the instantaneous Hamiltonian satisfies $H(\mathbf{K},t)\mathcal{I}=\mathcal{I}H(\mathbf{K},t)$, where $\mathcal{I}$ is the inversion operator.  For the present two-band model, this simply indicates that $H(\mathbf{K},t) = c_{\mathbf{K}}(t)\mathcal{I}$.  This greatly simplifies the time evolution operator at the HSMP because the Hamiltonian operator at different times commute with each other at HSMPs.   Therefore, $U_t(\mathbf{K})=\mathrm{exp}[-iC_{\mathbf{K}}(t)\mathcal{I}]$ where $C_{\mathbf{K}}(t)=\int_{0}^{t} c_{\mathbf{K}}(\tau)\,d\tau$. The inversion symmetry eigenvalues $\langle\sigma_x\rangle(\mathbf{K},t)\equiv\langle\phi(\mathbf{K},t)|\sigma_x|\phi(\mathbf{K},t)\rangle$ for the lower phase band $|\phi(\mathbf{K},t)\rangle$ are then fully determined by $C_{\mathbf{K}}(t)$, with
\begin{equation}\label{eq6}
  \langle\sigma_x\rangle(\mathbf{K},t)
  = -\mathrm{sgn}\Big(C_{\mathbf{K}}(t) \; \mathrm{mod}(-\pi,\pi]\Big).
\end{equation}

Now consider the two HSMPs $\Gamma$ and $M$, which are particularly important for understanding the topological phases at hand \cite{Zhu2021,Zhu2020}. The inversion symmetry eigenvalues along the time dimension, for the NI, AFI1, AFI2 and AFHOTI phases, are shown in Fig.~\ref{symmetry}.   At certain critical times, called SIPs, the inversion symmetry eigenvalue changes sign.  The NI does not exhibit any SIP [Fig.~\ref{symmetry}(a)], indicating that it can be continuously deformed into a normal insulator.  However, the two
AFI phases and the AFHOTI phase exhibit many SIPs [Fig.~\ref{symmetry}(b)-(d)].

Let us investigate the effective Hamiltonian $H_\mathrm{eff}(\mathbf{K},t)$ defined by $U_t(\mathbf{K})=\mathrm{exp}\big[-iH_\mathrm{eff}(\mathbf{K},t)\big]$, which describes time evolution up to time $t$.  There are two classes of SIPs, associated with the $\varepsilon=0$ and $\varepsilon=\pi$ gaps respectively. For the first class, the effective Hamiltonian in the neighborhood of a SIP is given by
\begin{equation}
  H_\mathrm{eff}(\mathbf{K},t)= \left.\frac{\partial C_{\mathbf{K}}(t)}{\partial t}\right|_{t_{\mathrm{SIP}}}\delta t \; \sigma_x,
\end{equation}
with $C_{\mathbf{K}}(t_{\mathrm{SIP}})=2n\pi$, $n\in \mathbb{Z}$.  For the second class,
the effective Hamiltonian around it is given by
\begin{equation}
  H_\mathrm{eff}(\mathbf{K},t) = \pi + \left.\frac{\partial C_{\mathbf{K}}(t)}{\partial t}\right|_{t_{\mathrm{SIP}}}\delta t\sigma_x
\end{equation}
with $C_{\mathbf{K}}(t_{\mathrm{SIP}})=(2n+1)\pi$, $n\in \mathbb{Z}$. The effective Hamiltonians for these two different classes of SIPs are marked in Figs.~\ref{symmetry}(b)-(d) by red (blue) for the $\Gamma$ ($M$) point.

For AFI1 [Fig.~\ref{symmetry}(b)], there are two SIPs at $\Gamma$: one for the $0$ gap and another for the $\pi$ gap. They are all stable because in the same gap, there is no partner to annihilate with.  For AFI2 [Fig.~\ref{symmetry}(c)], there is a single SIP at $\Gamma$ for the $\pi$ gap; this SIP is hence stable as well.  In addition, there are three SIPs at $M$ for the $0$ gap.  In this peculiar case, the middle SIP can annihilate with another (of the opposite eigenvalue) to its left or right, insofar as their movements in the time direction are not forbidden by particle-hole symmetry or inversion symmetry. This being the case, the three SIPs only contribute one stable SIP for the $0$ gap.  A similar analysis can be performed for the AFHOTI phase [Fig.~\ref{symmetry}(d)], yielding three SIPs at $\Gamma$; these are stable because one of them is for the $0$ gap, sandwiched between the other two for the $\pi$ gap.  The three SIPs at $M$, however, are all for the $0$ gap, and thus count as a single stable SIP.

Thus, an enhanced symmetry analysis that takes advantage of the detailed time dynamics can distinguish the AFI1, AFI2 and AFHOTI phases from each other, and from the other phases.   This is the main result of the present paper.

\begin{figure}
\includegraphics[width=1\linewidth]{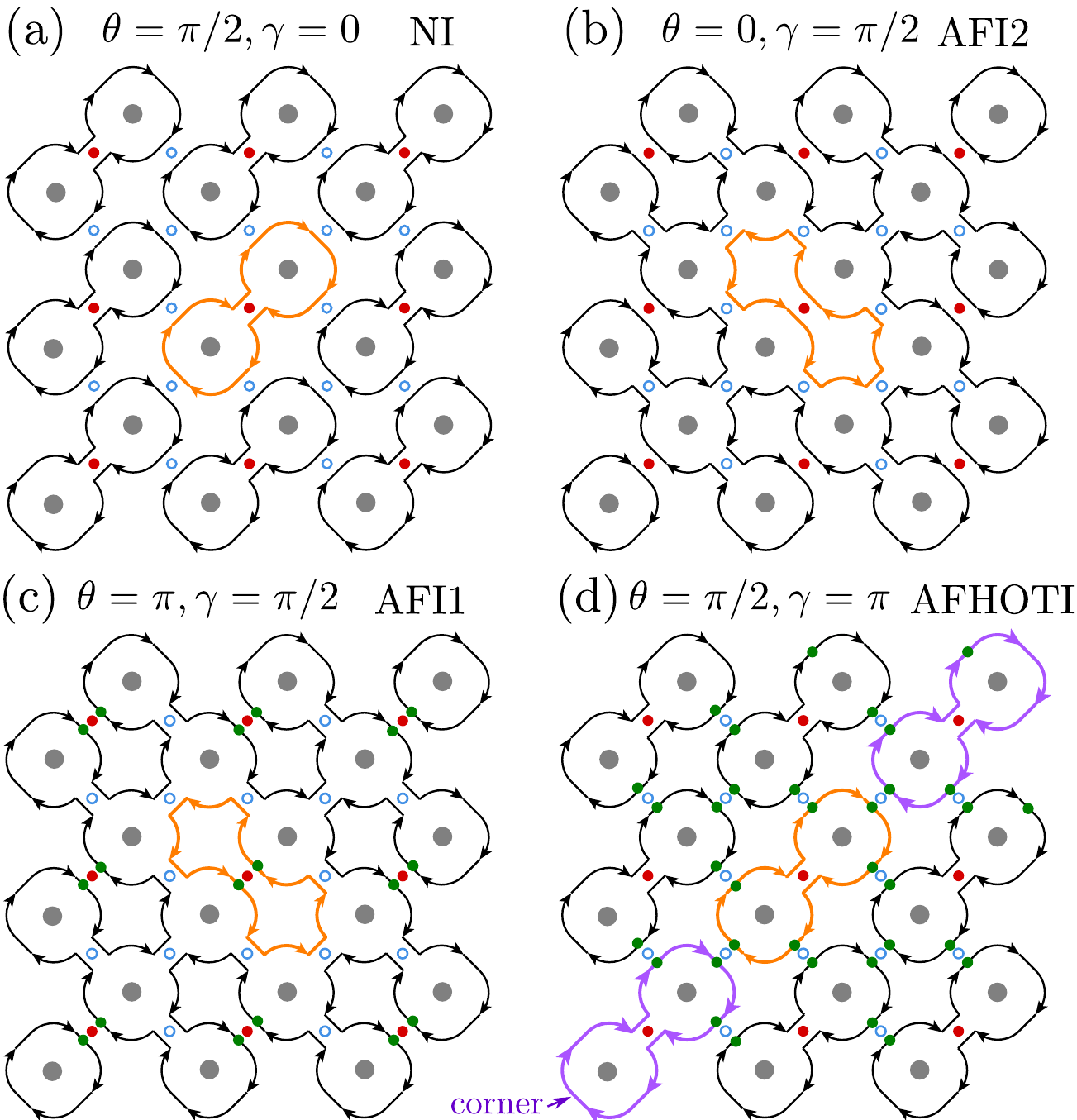}
\caption{Oriented scattering network representation for the four special cases in which a large network splits into decoupled pieces: (a) NI phase, (b)--(c) AFI phases, and (d) AFHOTI phase.  In each subplot, an exemplary localized bulk state is plotted in orange. Green dots indicate an additional $\pi$ phase shift during scattering.  The purple loop in (d) corresponds to a topological corner state.}
\label{localloop}
\end{figure}

{\it Wannier representation of anomalous Floquet topological states in a local loop formula.}---We have seen that AFI and AFHOTI phases cannot be distinguished from atomic insulators using Wannier obstruction, implying that their in-band bulk states can be deformed into localized states. Yet these phases are known to host topological chiral edge states (for the AFI) or topological corner modes (for the AFHOTI).  To understand why, we can turn to the oriented scattering network picture [Fig.~\ref{model}(b)], which is a way of representing Floquet lattices different from, but equivalent to, the usual description in terms of temporal modulation \cite{Liang2013}. Oriented scattering networks have previously been used to analyze integer quantum Hall~\cite{Chalker1988} and AFHOTI phases \cite{Zhu2021, Liu2020}.

In the oriented scattering network, the intra-cell scattering process $[c,~d]^T = S(\theta)~[a,~b]^T$ behaves specially when $\theta, \gamma \in \{0, \pi/2, \pi\}$.  At these values, $a$ ($b$) scatters entirely into $c$ ($d$) when $\theta=0$ or $\pi$, and $a$ ($b$) scatters entirely into $d$ ($c$) when $\theta=\pi/2$.  These special values of $\theta$ and $\gamma$ form nine distinct combinations. According to the phase diagram of Fig.~\ref{model}(c), only four of the combinations do not lie on phase boundaries: $(\theta,\gamma) \in \{(\pi/2,0), (0,\pi/2), (\pi,\pi/2), (\pi/2,\pi)\}$. These four combinations correspond to the NI, AFI2, AFI1 and AFHOTI phases respectively. The associated bulk quasienergy bands are $\varepsilon=\pm\pi/(2T)$, independent of momentum, meaning that the bulk states are fully localized.

As shown in Fig.~\ref{localloop}, the four special combinations cause the oriented scattering network breaks up into smaller decoupled pieces.  In the bulk, the wave amplitudes form localized loops, as shown by the exemplary orange paths in Fig.~\ref{localloop}(a)--(d).  The center of each local loop is the center of a unit cell, consistent with the Wyckoff positions being at $1a$. For the NI [Fig.~\ref{localloop}(a)], the loop encircles two sublattice centers (grey dots). For the AFIs [Fig.~\ref{localloop}(b) and (c)], the sublattice centers lie outside the loop; due to this configuration, it is possible to form a large loop encircling the entire finite lattice, which corresponds to the chiral edge states. For the AFHOTI phase [Fig.~\ref{localloop}(d)], the loop encircles the sublattice centers, similar to the NI case, and chiral edge states do not form.  However, certain loops localized to the lattice corners experience an odd multiple of $\pi$ phase shifts, which is qualitatively different from an edge local loop or a bulk local loop [e.g., the purple path in Fig.~\ref{localloop}(d)].  From this, we can deduce that topological corner states occur within the gaps at $\varepsilon = 0$ or $\varepsilon = \pi/T$.

{\it Conclusion and discussion.}---We have proposed a method to
conduct symmetry-based analysis and classification of Floquet topological phases.  The method is able to account for the rich variety of topological phases found in the two-dimensional square Floquet lattices.  In particular, stable singularities in the microscopic dynamics of Floquet topological systems can be identified by focusing on highly symmetric momentum points.  This allows for the characterization of Floquet topological phases that could not be distinguished from atomic insulators using standard analysis based on Wannier obstruction.  We have also applied this method to other lattice configurations, such as triangular lattices, where it identifies new topological states not found in the square lattice, as shown in the Supplemental Materials.

During the preparation of this manuscript, we noticed a preprint by Yu \textit{et al.}, which develops formal dynamical symmetry indicators for ``inherently dynamical'' Floquet crystals with first-order Floquet phases in  class A \cite{Yu2021}.

\begin{acknowledgements}
We acknowledges funding support by the Singapore Ministry of Education Academic Research Fund Tier-3 (Grant No. MOE2017-T3-1-001 and WBS No. R-144-000-425-592) and by the Singapore NRF Grant No. NRF-NRFI2017-04 (WBS No. R-144-000-378- 281).
\end{acknowledgements}

\clearpage
\onecolumngrid
\begin{center}
\textbf{\large Supplementary Materials}\end{center}
\setcounter{equation}{0}
\setcounter{figure}{0}
\setcounter{table}{0}
\renewcommand{\theequation}{S\arabic{equation}}
\renewcommand{\thefigure}{S\arabic{figure}}
\renewcommand{\thetable}{S\arabic{table}}
\renewcommand{\cite}[1]{\citep{#1}}

\section{Elementary band representations for square lattice with inversion symmetry}\label{App:band represnetation}

The elementary band representations (EBRs) for square lattice with inversion symmetry can be obtained by putting $s$ or $p$ orbital to maximal Wyckoff positions. There are eight EBRs. For completeness, all their inversion symmetry eigenvalues at HSMPs are shown in Table.~\ref{table1}. We note that each band representation has either 0, 2 or 4 negative eigenvalues.   
\\ \\ 
\begin{table}
\caption{\label{table1}Inversion symmetry eigenvalues at HSMPs for eight elementary band representations of square lattice with inversion symmetry. Each band representation has either 0, 2 or 4 negative eigenvalues. }
\begin{ruledtabular}
\begin{tabular}{ccccc}
& $\,\Gamma\,$ & $\,$ X$\,$& $\,$Y$\,$ & $\,$M$\,$ \\
\hline
  $s@q_{1a}$ & $+1$ & $+1$ & $+1$ & $+1$ \\
  $p@q_{1a}$ & $-1$ & $-1$ & $-1$ & $-1$ \\
  $s@q_{1b}$ & $+1$ & $-1$ & $+1$ & $-1$ \\
  $p@q_{1b}$ & $-1$ & $+1$ & $-1$ & $+1$ \\
  $s@q_{1c}$ & $+1$ & $+1$ & $-1$ & $-1$ \\
  $p@q_{1c}$ & $-1$ & $-1$ & $+1$ & $+1$ \\
  $s@q_{1d}$ & $+1$ & $-1$ & $-1$ & $+1$ \\
  $p@q_{1d}$ & $-1$ & $+1$ & $+1$ & $-1$ \\
\end{tabular}
\end{ruledtabular}
\end{table}
 
\section{A periodically driven triangle lattice with inversion symmetry}\label{App:band represnetation}

\begin{figure}[h]
\includegraphics[width=0.8\linewidth]{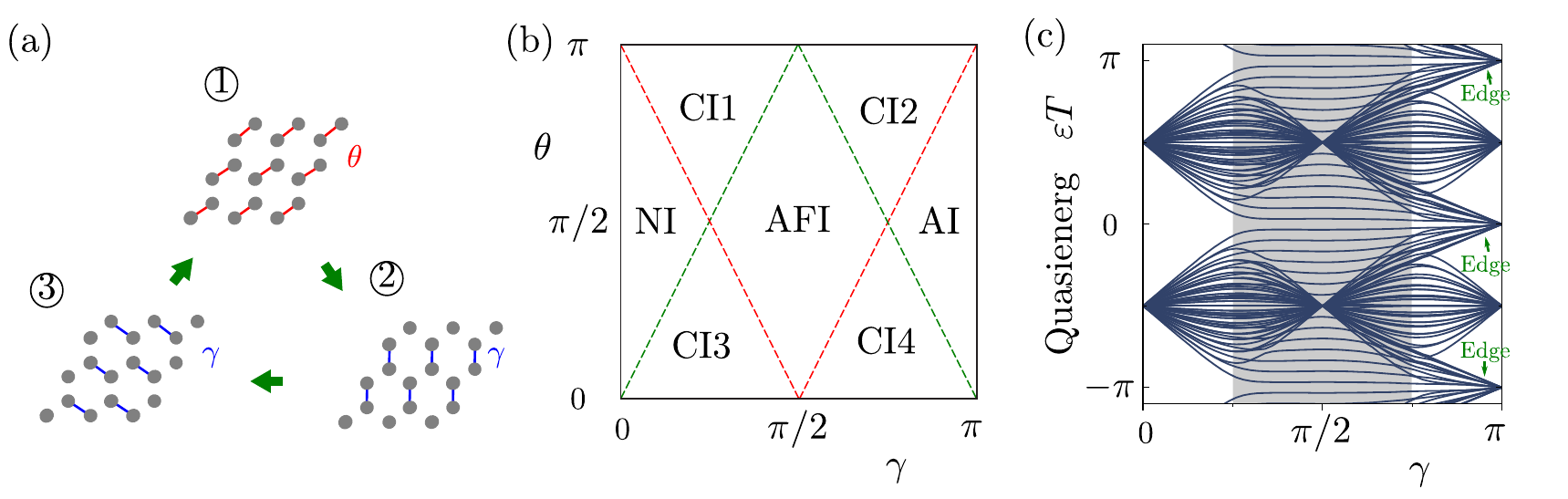}
\caption{Model and phase diagram in a periodically driven triangle lattice model. (a) Periodically driven bipartite model in the triangle lattice. The model is composed of three time steps. Step $\textcircled{1}$ contains intracellular couplings $\theta$. Step $\textcircled{2}$ and $\textcircled{3}$ contain intercellular couplings $\gamma$. (b) Phase diagram. The system supports different topological phases including one NI, one AFI, four CIs and one AI. (c) Spectrum of a finite structure as a function of $\gamma$ with $\theta=\pi/2$.}
\label{honeycomb}
\end{figure}

In this section, we carry out an analogous symmetry analysis in a periodically driving bipartite model in a triangle lattice as shown in Fig.~\ref{honeycomb}(a). The model is composed of three steps and can be described by the following time-dependent Bloch Hamiltonians,
\begin{eqnarray}
 H(\mathbf{k},t)=\left\{
\begin{array}{cc}
H_{1}(\mathbf{k})&0<t\leq T/3\;,\\
H_2(\mathbf{k})&T/3<t\leq 2T/3\;,\\
H_3(\mathbf{k})&2T/3<t\leq T\;,
\end{array}
\right.
\label{eqS1}
\end{eqnarray}
where,
\begin{equation}\label{eq2}
 H_{1}(\mathbf{k})=\theta\sigma_x\;,
\end{equation}
and,
\begin{equation}
H_{m}(\mathbf{k})=\gamma(e^{ib_m\cdot\mathbf{k}}\sigma^{+}+h.c.)\;,
\label{eq3}
\end{equation}
for $m=2,3$. $\sigma^{\pm}=(\sigma_{x}\pm i\sigma_{y})/2$, where $\sigma_{x,y,z}$ are Pauli matrices; and the vectors $\mathbf{b}_{m}$ are given by $\mathbf{b}_{2}=(a/2,\sqrt{3}a/2)$ and $\mathbf{b}_{3}=(a,0)$, where $a$ is the lattice constant. In our sample calculations, we set $T=3$.

Just like the square lattice in the main text, the triangle lattice also obeys particle-hole symmetry $CH(\mathbf{k},t)C=-H^{*}(-\mathbf{k},t)$ with $C=\sigma_z$ and inversion symmetry $\mathcal{I} H(\mathbf{k},t) \mathcal{I} = H(-\mathbf{k},t)$ symmetry, where $\mathcal{I}=\sigma_x$. The phase diagram is shown in Fig.~\ref{honeycomb}(b). There are different topological phases including one NI, four CIs, one AFI and one new phase as an anomalous insulator (AI). The existence of this particular AI will be explained later. Like NI, AI neither supports gapless chiral edge state nor supports topological corner modes. However, AI contains stable topological singularity in the microscopic dynamics and cannot be continuously deformed into NI (so we call it anomalous). In this way, AI is topologically distinguished from NI by the existence of stable SIPs as we show latter. AI is unique to Floquet systems and supports edge states gapped from bulk states. Fig.~\ref{honeycomb}(c) shows the spectrum of a finite structure as function of $\gamma$ with fixed $\theta=\pi/2$. From left to right, it is NI, AFI and AI. We note that the AI supports edge states gapped from bulk states.

\begin{table}
\caption{\label{table2}Inversion symmetry eigenvalues at HSMPs for eight elementary band representations of triangle lattice with inversion symmetry. Each band representation has either 0, 2 or 4 negative eigenvalues. }
\begin{ruledtabular}
\begin{tabular}{ccccc}
& $\,\Gamma\,$ & $\,$ M$_1$$\,$& $\,$M$_2$$\,$ & $\,$M$_3$$\,$ \\
\hline
  $s@q_{1a}$ & $+1$ & $+1$ & $+1$ & $+1$ \\
  $p@q_{1a}$ & $-1$ & $-1$ & $-1$ & $-1$ \\
  $s@q_{1b}$ & $+1$ & $-1$ & $+1$ & $-1$ \\
  $p@q_{1b}$ & $-1$ & $+1$ & $-1$ & $+1$ \\
  $s@q_{1c}$ & $+1$ & $-1$ & $-1$ & $+1$ \\
  $p@q_{1c}$ & $-1$ & $+1$ & $+1$ & $-1$ \\
  $s@q_{1d}$ & $+1$ & $+1$ & $-1$ & $-1$ \\
  $p@q_{1d}$ & $-1$ & $-1$ & $+1$ & $+1$ \\
\end{tabular}
\end{ruledtabular}
\end{table}
\begin{figure}[h]
\includegraphics[width=0.7\linewidth]{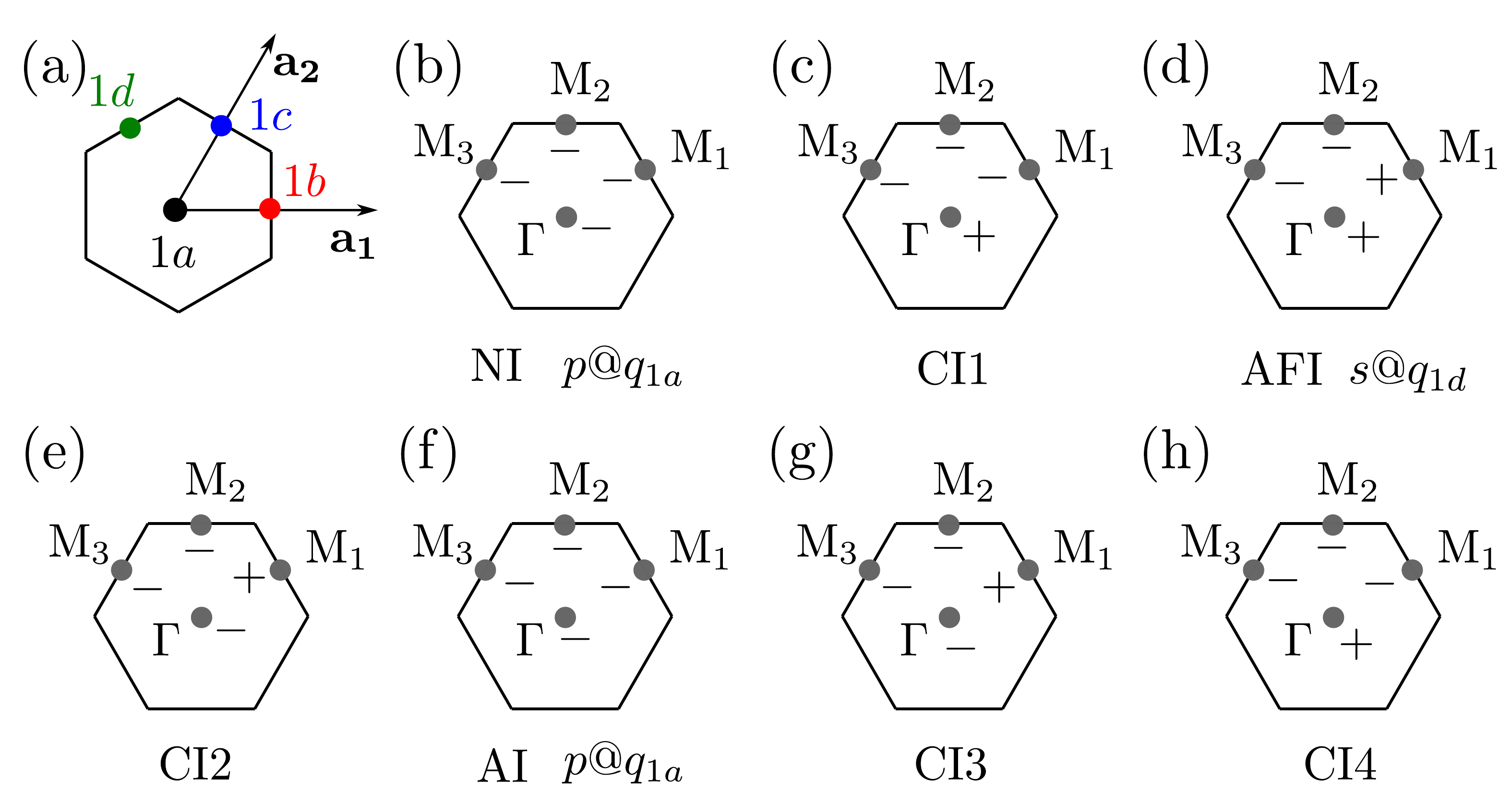}
\caption{Maximal Wyckoff positions and inversion symmetry eigenvalues for our triangle lattice model. (a) Maximal Wyckoff positions of the triangle lattice with inversion symmetry. (b)-(h) Inversion symmetry eigenvalues at HSMPs for different topological phases. $\Gamma$, M$_1$, M$_2$ and M$_3$ are high symmetric points in the Brillouin zone.}
\label{Wyckofftriangle}
\end{figure}

Next, we carry out a general analysis for the triangle lattice system here with inversion symmetry and provide the EBRs. The unit cell of the triangle lattice is shown in Fig.~\ref{Wyckofftriangle}(a). There are four maximal Wyckoff positions $q_{1a}$, $q_{1b}$, $q_{1c}$ and $q_{1d}$. By putting $s$ or $p$ orbital to those points, we can obtain eight EBRs. The inversion symmetry eigenvalues at HSMP for those eight EBRs are summarized in Table.~\ref{table2}.

We then calculate the inversion symmetry eigenvalues for different topological phases in Fig.~\ref{honeycomb}(b). The results are shown in Figs.~\ref{Wyckofftriangle}(b)-(h). By comparing with the EBRs in Table.~\ref{table2}. We see that NI, AFI and AI have corresponding EBRs $p@q_{1a}$, $s@q_{1d}$ and $p@q_{1a}$, which means that their band Chern number is zero. The four CIs do not have EBRs.

\begin{figure}[h]
\includegraphics[width=0.8\linewidth]{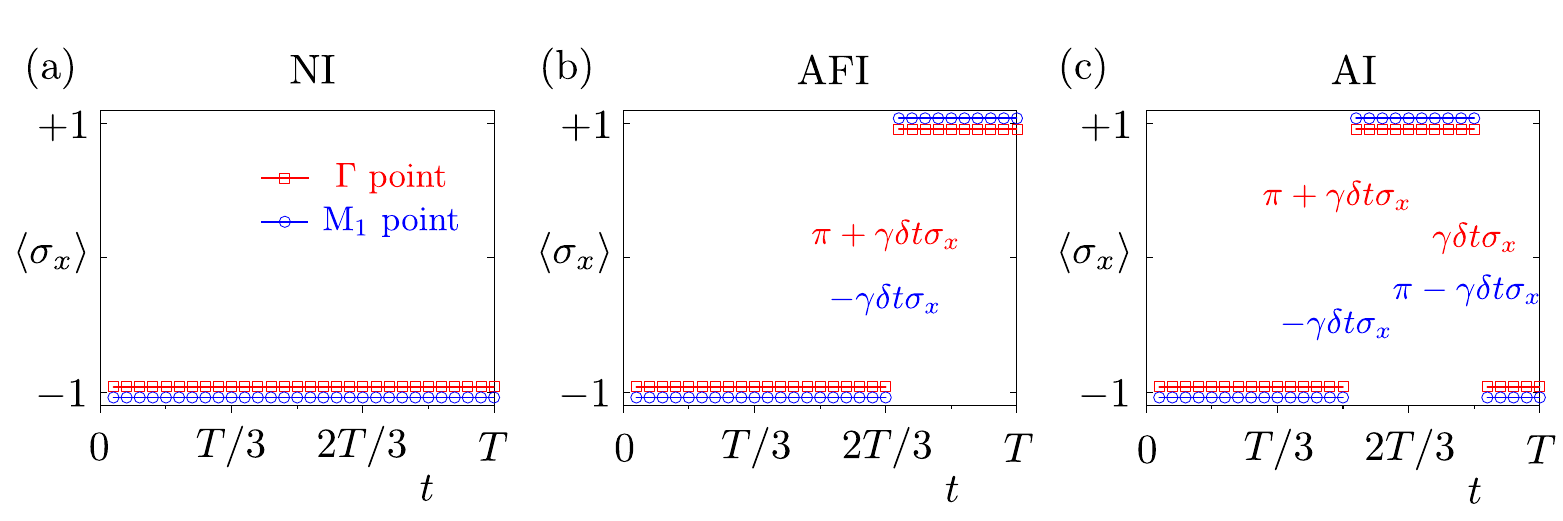}
\caption{Stable SIPs along the time dimension in microscopic dynamics in our driven triangle lattice model (a)-(c) are results for NI, AFI and AI. The continuous effective Hamiltonian around the SIPs are marked in different colors as the main text. In our calculations, we choose $\theta=\pi/2$, $\gamma=0$ for (a), $\theta=\pi/2$, $\gamma=\pi/2$ for (b) and $\theta=\pi/2$, $\gamma=\pi$ for (c).}
\label{SIP}
\end{figure}

Next we study the symmetry eigenvalues of the microscopic time evolution operator for NI, AFI and AI. The results for $\Gamma$ and M$_1$ points are shown in Fig.~\ref{SIP}. It is seen that NI does not have SIPs, AFI has one stable SIP in each quasienrgy band gap (0 and $\pi$), and AI contains two stable SIPs in each quasienergy band gap.

We highlight that the sole symmetry analysis described above may not completely distinguish between different topological phases. If the system supports odd number SIPs in one gap, we can confirm that it supports gapless chiral edge states. If the system does not support SIP, we can confirm that it is NI. These two situations are clear cut cases. But if the system contains nonzero and an even number of SIPs, it may support even gapless chiral edge states, topological corner states (AFHOTI) or edge states gapped from bulk state (AI). They can be further distinguished by calculating other necessary topological invariants involving the whole momentum space.  However the existence of stable SIP promises that the system must be topologically different from NI.  As such, our easy-to-implement method can be used to quickly detect topologically nontrivial states in Floquet topological systems.

\end{document}